# Capturing Differences in Character Representations Between Communities: An Initial Study with Fandom

Bianca N.Y. Kang[1][0009-0007-2745-0529]

[1] Carnegie Mellon University, Pittsburgh PA 15213, USA
`biancak@andrew.cmu.edu`

**Abstract.** Sociolinguistic theories have highlighted how narratives are often retold, co-constructed and reconceptualized in collaborative settings. This working paper focuses on the re-interpretation of characters, an integral part of the narrative story-world, and attempts to study how this may be computationally compared between online communities. Using online fandom – a highly communal phenomenon that has been largely studied qualitatively – as data, computational methods were applied to explore shifts in character representations between two communities and the original text. Specifically, text from the Harry Potter novels, r/HarryPotter subreddit, and fanfiction on Archive of Our Own were analyzed for changes in character mentions, centrality measures from co-occurrence networks, and semantic associations. While fandom elevates secondary characters as found in past work, the two fan communities prioritize different subsets of characters. Word embedding tests reveal starkly different associations of the same characters between communities on the gendered concepts of femininity/masculinity, cruelty, and beauty. Furthermore, fanfiction descriptions of a male character analyzed between romance pairings scored higher for feminine-coded characteristics in male-male romance, matching past qualitative theorizing. The results highlight the potential for computational methods to assist in capturing the re-conceptualization of narrative elements across communities and in supporting qualitative research on fandom.

## 1 Introduction

### 1.1 Collaboration and narratives

Narratives are a medium for signaling their tellers' identities, values, and perspectives [1]. Current sociolinguistic theories point towards narrative as a highly interpretative and collaborative exercise [1], [2], [3]. Social groups frequently engage in collaborative storytelling [2], [4], which helps reinforce group membership [1]. These collaborative acts also underscore another feature of narratives – their fluidity. Instead of existing as complete, unmalleable units, they are often in flux, open to change and renegotiation across interactions and contexts [2]. People, relationships, events, and values are often reconceptualized in story-telling [4].

The structure and affordances of today's social media platforms – comments, forum discussions, episodic updates – provide rich contexts to further our understanding of narrative non-linearity and collaborative construction and re-construction, beyond prior offline ethnographic studies [3], [5], [6]. In their networks, users remix and remake content from others [5]; the role of the community takes centerstage online [3]. The availability of online data and computational techniques (e.g., in framing) highlight a potential to study at scale how components of a story-world (characters, situations, events [1]) might be re-negotiated across communities. *Characters* and their relationships are central to modern analyses of narratives [7], our understanding of society [8], and phenomena like conspiracies [9]. In this working paper, I thus specifically explore how changes in *character* representations might be captured quantitatively as they are retold in different online communities with varying communicative contexts and goals.

### 1.2   Online fandom

Drawing from recent studies that have recognized online fandom as a resource for understanding broader phenomena like re-interpretation [10] and misinformation [11], this exploration is conducted via Harry Potter fan communities, which have had significant influence on modern Internet culture [12]. Specifically, character representations are compared between a fanfiction (fan-written prose based on some source material) community, fan subreddit, and canon (source material, taken as the original 7 books).

Fandom is highly communal [13] and social [14], making it ideal for studying collaborative narrative reconstruction. Through fanfiction writings, interactions between fanfiction readers and writers, and discussions on forums, modified or even new plot and character concepts can arise and be established within communities as fanon [3], [15]. Though fanfiction and forums have different communicative contexts (prose vs. messages), these online interactions are akin to each other, and offline co-construction of personal narratives, in their collaborative nature. Furthermore, offline narrative interactions are attempts to make sense and provide subjective *perspectives* of experiences [4]. Similarly, fandom participation often involves analysis, re-working, and re-telling of characters and events [10], [14], [16], though specific motivations for doing so are diverse. For instance, fanfiction may attempt to cover perceived gaps in representations in canon [16], while forum discussions may relate characters and values from canon back to real life [17].

A large body of qualitative work has examined how fandom reworks narratives from canon. Computational studies are growing in the area [10], [18], [19]; methods like mention counts, embeddings, and descriptions have been used to capture representation shifts between *fanfiction* and canon. This paper extends such work by including (1) a fan forum, to investigate representation differences *between communities* and (2) an exploration on using semantic axes to capture changes in stereotypical attributes, following recent computational research on framing. Additionally, it contributes to the limited body of work analyzing fanfiction stories quantitatively [14].

### 1.3 Research questions

This paper takes from the mental model of a story-world ("*who* did *what* with *whom*…in *what manner*" [1, p.122]). I examine how characters shift in (1) space allocated (*who*), (2) centrality measures (*with whom*), and (3) semantic associations between the source text and two online communities (*what*, and *in what manner*).

Space allocation is quantified by mention counts and taken as a measure of attention allocated [18], [20]. Centrality measures are derived from networks constructed based on the common definition of simple character co-occurrence counts [21]. For each source, a single network is created by combining all its texts. While a fanfiction story or Reddit comment can be read as an isolated textual unit, the design of online platforms and user browsing behaviors mean readers typically move across content freely, creating associations and broader contexts between non-adjacent units [3], [6]. Weighted degree centrality (the overall strength of ties) and core-periphery scores (if the character is a densely connected core, or peripheral) [22] are used to evaluate a character's connection density. Betweenness centrality (bridging), closeness centrality (closeness to other characters), and effective size (neighborhood redundancy) are used to evaluate how well a character links other characters.

Semantic axes are used to evaluate semantic associations, following research using them to evaluate social group stereotypes (e.g., [20], [23]). In a large-scale study of Wattpad fanfiction, Fast et al. found certain descriptive categories associated more with male than female characters (e.g., beauty, violence) [24]. The axes of *masculine-feminine*, *ugly-beauty*, *cruel-kind* (seed words from [25]) were thus selected for analysis. Given the dominance of shipping (i.e., romantic pairings) subcommunities in fanfiction [10], [12], an exploration was also conducted on how descriptions assigned to a character might change between different romantic contexts. Attributed verbs, adjectives, and nouns were extracted [20], [24] and subjected to semantic axis tests.

## 2 Methods

### 2.1 Data collection

The Harry Potter novels were taken from Github[1]. 48952 posts and 661602 comments posted on r/HarryPotter from January 1 2020 to January 1 2021 were collected with PMAW and PRAW. 35836 Harry Potter fanfiction written in English and published between January 1 2020 to January 1 2021 were scraped from Archive of Our Own (AO3).

### 2.2 Data preparation

For mention counts, the novels were processed with the large BookNLP model [26], which performs character name clustering (e.g., 'Mr. Potter' to 'Harry Potter'). Only proper nouns were used with some manual updating (e.g., 'Padfoot' to 'Sirius Black')[2]. For r/HarryPotter, post titles and text were combined, and named entity recognition (NER) with SpaCy identified character names. Well-known permutations of the top-50 names were aggregated (e.g., 'remus lupin', 'moony' to Remus Lupin). The process was replicated for the AO3 dataset.

---

[1] https://github.com/idc9/stor390/tree/master/notes/natural_language_processing/rowling

To create a common character set for network analysis, the top 100 characters mentioned in r/HarryPotter and AO3 were identified, resulting in a list of 123 characters. Canon was not used to avoid missing secondary characters discussed more by fans. As canon and AO3 contained long-form stories, their networks were created the same way. Characters in the same paragraph were linked, with the link weighted by the number of paragraphs they co-occurred in. The individual novel networks were combined to one, and same with AO3. The r/HarryPotter network was created differently due to Reddit's thread structure. Characters were linked if they appeared in the same comment tree, with the link weighted by the number of shared comment trees.

For semantic associations, following [27], separate word2vec models were trained for each source, using the Gensim skip-gram model with default parameters. CoreNLP and a set of dependency parsing heuristics were used to extract descriptions (nouns, adjectives, and subject-verbs). Verbs were lemmatized, negated descriptions removed.

## 3      Results and Discussion

### 3.1      Shifts in space allocation

The top mentioned characters varied across source (Table 1). A chi-square test of independence was done to assess if the communities allocated attention differently to characters from canon and each other. There was a significant ($p<.001$), $\chi 2(24) = 259033.152$ and moderate (Cramer's V = .209) association between mention count and source. Consistent with prior research, the table shows female characters receiving increased attention in fandom versus canon [18].

Table 1. Top 10 mentioned characters from canon and fan communities; (F) – female.

|    | Canon               | r/HarryPotter        | AO3                  |
|----|---------------------|----------------------|----------------------|
| 1  | Harry Potter        | Harry Potter         | Harry Potter         |
| 2  | Ron Weasley         | Lord Voldemort       | Draco Malfoy         |
| 3  | Hermione Granger (F)| Severus Snape        | Hermione Granger (F) |
| 4  | Albus Dumbledore    | Albus Dumbledore     | Ron Weasley          |
| 5  | Lord Voldemort      | Hermione Granger (F) | Severus Snape        |
| 6  | Rubeus Hagrid       | Ron Weasley          | Sirius Black         |
| 7  | Severus Snape       | Sirius Black         | Lord Voldemort       |
| 8  | Sirius Black        | Draco Malfoy         | Ginny Weasley (F)    |
| 9  | Draco Malfoy        | James Potter         | Lily Potter (F)      |
| 10 | Fred Weasley        | Lily Potter (F)      | James Potter         |

---

[2] BookNLP performs co-reference resolution, but those were not used due to errors (e.g., pronouns for 'Ron' identified wrongly as 'Harry').

### 3.2  Shifts in centrality measures

Centrality measures were calculated with NetworkX. Core-periphery scores were calculated following [22]. The measures were converted to ranks to compare across sources, with 1 as the highest scorer. For measures relating to connection density (weighted degree, core-periphery score; Figure 1), characters who score highly in canon (main characters like *Harry*) also score highly in fan writings. Some less densely-connected characters have large improvements in degree and core position in fan writings, consistent with past work showing secondary characters have increased presence in fanfiction [18].

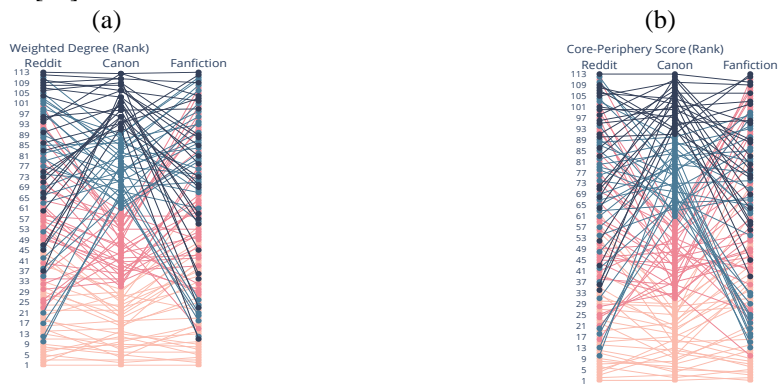

**Fig. 1.** Shifts in character rankings in (a) weighted degree and (b) core-periphery score from r/HarryPotter, canon, to AO3. Lighter colors = higher ranks in canon.

For measures relevant to how well the character *links others* (betweenness, closeness, effective size; Figure 2), characters who link others well in canon (again, typically main characters) do not maintain this quality in fan writings, particularly in fanfiction. In both communities, a subset of characters cluster onto the same low ranks for betweenness and closeness. For characters with high effective size in canon, their ego networks become redundant (i.e., neighbors are now well-connected to each other).

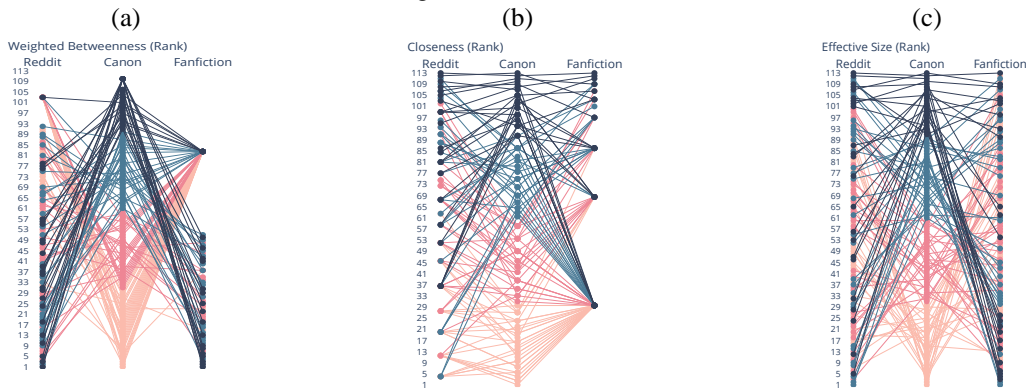

**Fig. 2.** Shifts in character rankings in (a) weighted betweenness, (b) closeness, and (c) effective size from r/HarryPotter, canon, to AO3. Lighter colors = higher ranks in canon.

Both observations suggest that fan communities, particularly fanfiction, focus their attention on the relationships of a limited set of characters – canon main characters and a select subset of secondary ones. These results add nuance to past findings of fanfiction uplifting secondary characters [18], revealing that not all secondary characters receive the same boost in attention. In addition, these observations offer a starting point for further work in identifying the attributes of characters that garner more attention within and across communities.

### 3.3    Shifts in semantic associations

The canon and r/HarryPotter word2vec models, and the canon and AO3 models, were aligned with a Procrustes transformation [28]. Pairwise cosine similarity for 57 commonly-mentioned character names was calculated for both model pairs (Figure 3). Neither community is consistently higher in name semantic similarity to canon. As with the preceding sections, the extent of change in semantic context from canon varies depending on the character and community in question.

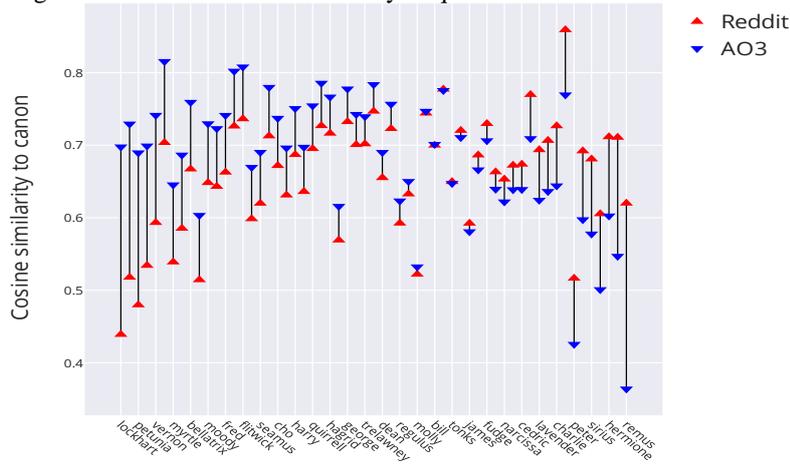

**Fig. 3.** Pairwise cosine similarity of character names between r/HarryPotter and canon, and AO3 and canon. Only some names shown for clarity.

Semantic axis tests (Relative Norm Distance [29] for *masculine-feminine*, SEMAXIS for the antonym tests of *cruel-kind* and *ugly-beauty* [30]) were conducted with character names using the r/HarryPotter and AO3 models separately[3]. Characters were ranked by their cosine similarity score for each model, with 1 being the closest scorer to the left pole (*masculine*, *cruel*, *ugly*) (Figure 4).

---

[3] The pairwise cosine similarities of the seed words on each axis pole between r/HarryPotter and AO3 models were evaluated with aligned models. The average similarity ranged from .664 to .752. As a baseline, average pairwise stop-word cosine similarity was .721.

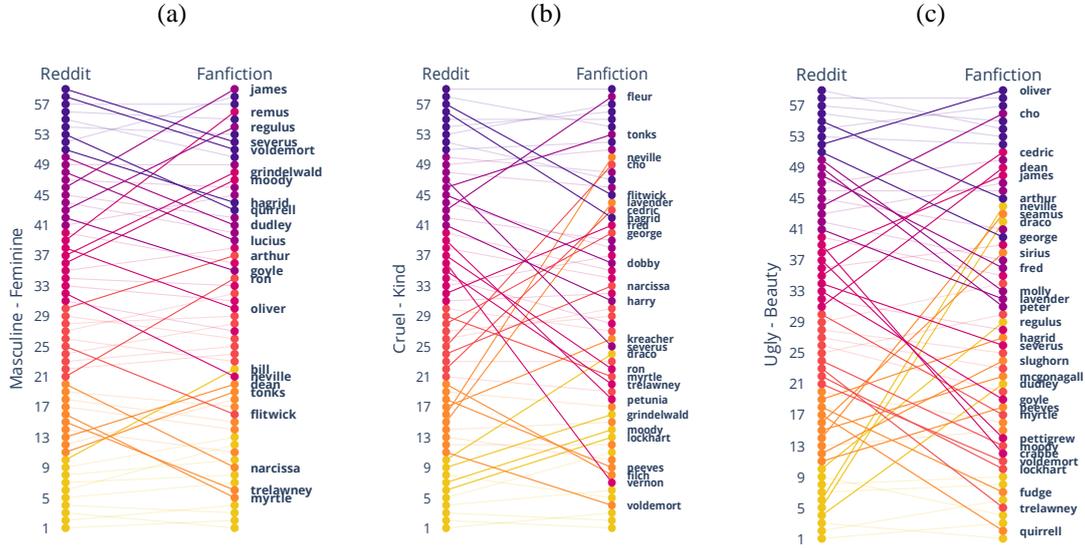

**Fig. 4.** Shifts in ranks on (a) Masculine – Feminine, (b) Cruel – Kind, (c) Ugly – Beauty between r/HarryPotter and AO3. Shifts greater than 6 ranks are opaque. Lighter colors = higher ranks on r/HarryPotter.

The rank shifts highlight stark differences in the semantic associations of character names between the two fan communities, despite their shared fandom. While there is some agreement at the polar ends, there are overall large divergences for most characters A notable example: *Draco* shifts 14 ranks towards *kindness* and 40 ranks towards *beauty* in AO3.

For the final exploration within AO3 fanfiction, *Draco* was selected due to the large differences in their semantic associations and space allocation between communities. I investigated their attributed descriptions in the varying context of a homosexual (slash) and heterosexual *ship*[4]. The introduction of feminine-coded characteristics to male characters in slash has long been discussed qualitatively [31]. Two strongly gender-coded categories were thus selected from [24] for the axes – *submissive-dominant*, *weak-strong*. Descriptions with a weighted log-odds ratio score [32] greater than 1.64 were used for semantic axis tests. On average, descriptions of *Draco* in slash are more associated with weakness (*mean*= -.15, *SD*=.14) and submissiveness (*mean*=-.18, *SD*=.13) than in the heterosexual context (*mean*= -.02, *SD*=.11) and submissiveness (*mean*=-.06, *SD*=.1), providing evidence for past qualitative observations.

---

[4] The two relationships investigated are the character's most popular in the AO3 dataset (with Harry, and with Hermione).

## 4      Conclusion

This working paper examined the usage of computational methods to analyze differences in character representations constructed by two online communities. Both communities allocate more space to secondary and female characters, consistent with past research. They also have different semantic associations for characters; even within fanfiction, differences arise between shipping sub-communities. Writings of a male character in the slash context use more feminine-coded descriptions versus the heterosexual context, matching past qualitative observations. Centrality measures show the attention boost for secondary characters is focused on a select subset. This suggests avenues for further exploration, like the identification of character attributes linked to larger shifts in attention and semantic association. These differences may have significance; character roles may be shaped by external movements [10], stereotyping [4], or qualities (e.g., hyperdiegesis [13]).

An extension would be to apply these methods to other online communities also characterized by extensive co-construction of narratives, like conspiracy groups. Computational analysis of conspiracies as actant networks is growing [9], [33]. Integrating more multi-faceted evaluations of character representations shifts (especially semantically) over time and across communities could help us – extending from [4]'s work on conversational personal narratives – get a better understanding of how different groups use reconceptualization to navigate their pasts, presents, and negotiate possible futures.

In future research, I would more explicitly consider how the affordances and presentation of information on platforms can affect reconstruction [5], [7] and weigh the influence of different user roles [5]. While narratives are a collaborative process, the credibility and power held by different tellers influence which representations are kept [4]; fandom too has social hierarchies [13]. Methods-wise, I would evaluate more robust approaches for character network creation, given the arbitrariness of paragraph splitting and non-specificity of co-occurrence [21]. I would also explore the potential of contextualized embeddings for semantic axis testing [34] and character embeddings [35].